\documentclass[12pt]{article}
\usepackage{graphicx}
\usepackage{subfigure}
\usepackage{amssymb}
\usepackage{amsfonts}
\usepackage{latexsym}
\usepackage[dvips]{color}

\input epsf.tex

\newcommand{\beq}{\begin{equation}}
\newcommand{\eeq}{\end{equation}}
\newcommand{\be}{\begin{equation}}
\newcommand{\ee}{\end{equation}}
\newcommand{\bea}{\begin{eqnarray}}
\newcommand{\eea}{\end{eqnarray}}

\makeatletter
\renewcommand{\theequation}{\thesection.\arabic{equation}}
\@addtoreset{equation}{section}
\makeatother

\newcommand{\tr}{\mathop{\rm tr}}

\def\href#1#2{#2}

\def\bra#1{| #1 \rangle}
\def\ket#1{\langle  #1 |}

\begin{document}

\baselineskip=15.5pt
\pagestyle{plain}
\setcounter{page}{1}

\begin{titlepage}
\begin{flushleft}
       \hfill                       FIT HE - 20-01 \\
       \hfill                       
\end{flushleft}

\begin{center}
  {\huge Extension to Imaginary Chemical Potential \\   
   \vspace*{2mm}
in a Holographic Model 
\vspace*{2mm}
}
\end{center}

\begin{center}

\vspace*{5mm}
{\large ${}^{\dagger}$Kazuo Ghoroku\footnote[1]{\tt gouroku@fit.ac.jp},
${}^{\dagger}$Kouji Kashiwa\footnote[2]{\tt kashiwa@fit.ac.jp},
Yoshimasa Nakano\footnote[3]{\tt ynakano@kyudai.jp},\\
${}^{\S}$Motoi Tachibana\footnote[4]{\tt motoi@cc.saga-u.ac.jp}
and ${}^{\ddagger}$Fumihiko Toyoda\footnote[5]{\tt f1toyoda@jcom.home.ne.jp}
}\\

\vspace*{2mm}
{${}^{\dagger}$Fukuoka Institute of Technology, Wajiro, 
Fukuoka 811-0295, Japan\\}
\vspace*{2mm}
{${}^{\S}$Department of Physics, Saga University, Saga 840-8502, Japan\\}
\vspace*{2mm}
{${}^{\ddagger}$Faculty of Humanity-Oriented Science and
Engineering, Kinki University,\\ Iizuka 820-8555, Japan}
\vspace*{3mm}
\end{center}

\begin{center}
{\large Abstract}
\end{center}

We extend a bottom up holographic model, which has been used in studying the color superconductivity in QCD,  
to the imaginary chemical potential ($\mu_I$) region, and 
the phase diagram is studied on the $\mu_I$-temperature (T) plane.
The analysis is performed for the case of the probe approximation
and for the background where the back reaction from the flavor fermions are taken into account. 
For both cases,  
we could find the expected Roberge-Weiss (RW) transitions. 
In the case of the back-reacted solution, a bound of the color number $N_c$ is found to produce the
RW periodicity. It is given as $N_c\geq 1.2$. 
Furthermore, we could assure the validity of this extended model 
by comparing our result with
the one of the lattice QCD near $\mu_I=0$.

\noindent

\begin{flushleft}

\end{flushleft}
\end{titlepage}

\vspace{1cm}

\section{Introduction}

The holographic approach based on the string/gauge duality 
is a powerful method to study various thermo-dynamical 
and non-perturbative properties of the Yang-Mills theory
especially when the chemical potential ($\mu$) of the fundamental fermions plays an important role. 
Various results have been
obtained in such  cases and they could give us many kinds of insight into the phase diagram of QCD. (See for example
\cite{mateos2,Horigome,GKTT,GKTTT}.)

On the other hand, many non-perturbative
investigations in QCD have been performed by the lattice gauge theory. However, the analysis has been restricted
to the case of the imaginary chemical potential, $\mu_I$, to avoid the sign problem of the fermion determinant
(See for example \cite{Bilgici:2008qy,Bilgici:2009tx}). 
In QCD with $\mu_I$, on the other hand, the Roberge-Weiss (RW) phase transition
and its periodicity with respect to $\mu_I$ have been pointed out
as a remarkable point 
\cite{rw}. This observation is understood from the periodicity of the partition function.
It would be meaningful to see how
this point is realized in the holographic approach 
to make clear the validity of the holographic approach
.

Ten years ago, however, such a holographic investigation 
has been given 
based on the Euclidean space-time geometry \cite{Aarts,Raff,Morita}.
In the Ref. \cite{Aarts} 
, the periodic RW transition 
has been shown 
by adding 
the two-form Kalb-Ramond field $B$ 
in the D3/D7-brane system of the type IIB model.
After that, also in the 
D4/D8-brane system in IIA model, similar analysis has been done in Ref. \cite{Raff}, and also in Ref. \cite{Bigazzi:2014qsa, Morita} in a slightly different method. 
In these approaches, 
the essential point is the introduction of the $B$ field 
with dB=0 and its potential $V_{A}(\alpha)$, where 
$\alpha$
corresponds to the phase of the Polyakov loop \cite{witten1,Aharony-w}. 
It is introduced as
\footnote{The disc $D_2$ is a part of the black hole geometry considered here.}
\beq\label{B-alpha}
\alpha = \int_{D_2} {B\over 2\pi\alpha'}\, .
\eeq
And this parameter $\alpha$ discriminates the periodic vacua in the deconfinement phase with spontaneously broken $Z_N$ symmetry. 
On the other hand, 
$\mu_I$ comes from the bulk U(1) gauge field $F(=F_{\mu\nu}dx^{\mu}\wedge dx^{\nu})$, and it appears in the theory being combined
with $\alpha$ as
\beq\label{Bfield}
  \alpha-{\mu_I \over T} = \int_{D_2}\left(F+ {B\over 2\pi\alpha'}\right)\, .
\eeq

The  
important point is that the potential $V_{A}(\alpha)$
is periodic under $\alpha \to \alpha +2\pi /N_c$ 
due to the gauge symmetry of the boundary SYM theory \cite{rw} 
\footnote{See also the appendix A.}.
As a result, the total effective potential, the sum of $V_{A}(\alpha)$ and the probe action, is also periodic under $\mu_I/T \to \mu/T +2\pi /N_c$,
since a finite shift of $\mu_I/T$ can be absorved into $\alpha$ as understood from (\ref{Bfield}). 
The role of $\mu_I$ in the probe action is to 
control the minimum of the effective total potential of $\alpha$ as seen in the RW phase transition \cite{rw, Aarts}.

The purpose of this paper is to extend the analysis performed for $\mu_I$ 
in the top down models to a bottom up model which has been
used to study the color superconductivity in QCD \cite{Gkntt,Fadafan,Basu}. 
Through the extension, we could get phase diagrams for our model in the region of $\mu_I$ with RW transitions.  
And 
an implication of our holographic model is discussed.
Especially, when the back-reaction of flavor fermions is included, we find a $\mu_I$-dependent
critical curve of confinement/deconfinement transition. 
In this case,
we could show the usefulness of 
a simple continuation, $\mu\to i \mu_I$, in terms of the critical curve obtained 
for real $\mu$. 
This usefulness is supported by the fact that we can set as $\alpha =0$ near $\mu_I=0$.  

\vspace{.3cm}
 In the next section,
the extended bottom up model is proposed and the actions are estimated for confinement and deconfinement phases. In Sec.3, 
the RW transitions are investigated in the probe approximation, and for the back reacted case. 
Then 
the phase diagrams are given. In Sec. 4, the validity of the continuation near $\mu=0$ is discussed 
by comparing the critical curve
near $\mu=0$ for the holographic model and the one of the dual QCD theories. A problem related to a wide periodicity
of the potential of $\mu_I$ is discussed in Sec.5.
Our summary is given in the final section.

\vspace{.3cm}

\vspace{.3cm}

\section{A bottom up model} 

A bottom up model, which is used before  to study the superconductivity of QCD \cite{Gkntt}, is 
given in a slightly modified form of the following action for the Euclidean space-time
to investigate QCD with the imaginary chemical potential.
\beq
S = S_{\rm bu}+S_{F_{(4)}}\,,
\eeq
where the first term is given as
\bea\label{bottom-up}
S_{\rm bu} &=& \int d^{6} x \sqrt{-g}\, \left({\cal L}_\mathrm{Gravity}+{\cal L}_\mathrm{CSC} \right) \,,  \\
     {\cal L}_\mathrm{Gravity}&=& {1\over 2\kappa_6^2}\left( {\cal R} + {20 \over L^2}\right)\,, \label{bulk-L} \\
  \tilde{ {\cal L}}_\mathrm{CSC} &=& - {1 \over 4} \tilde{F}^2 - |D_{_\mu} \psi|^2 - m^2 |\psi|^2\, , \label{probe-L} \\
  \tilde{F}_{\mu\nu}&=& 
\partial_\mu A_\nu-\partial_\nu A_\mu+{B_{\mu\nu}\over 2\pi\alpha'}\,,\quad
D_{\mu} \psi = (\partial_{\mu}-iqA_{\mu})\psi\, .
\eea
This action $S_{\rm bu}$ is proposed as a model dual to the SYM theory with strongly interacting flavor fermions with the chemical potential $\mu$
when the space-time is Lorentzian and $B_{\mu\nu}$ is neglected. In the gravitational action ${\cal L}_\mathrm{Gravity}$, the scale 
$L$ denotes the AdS radius.

In the present case, the theory is set
in the Euclidean space-time. It is obtained by the Wick rotation of both the time and fields. Furthermore, the Kalb-Ramond
field is added through $\tilde{F}_{\mu\nu}$. 
This form of $\tilde{F}_{\mu\nu}$ is implied from the D-brane action.  
And
$\psi$ denotes a charged scalar, which is supposed to be dual to the cooper pair of the color charged fermions. Its baryon number charge is assigned as $q$. 
We could show that
there is no non-trivial solution for $\psi$ in the region of small and negative $\mu^2$ \cite{Gkntt}. 
Since $\mu$ is imaginary for negative $\mu^2$, we can neglect $\psi$ hereafter because we are considering 
the case with the imaginary chemical potential.

Then, the system can be solved by setting as $\tilde{A}_0=\tilde{\phi}$
where $\tilde{A}_\mu$ is defined by
\beq
   \tilde{F}_{\mu\nu}  \equiv \partial_\mu \tilde{A}_\nu-\partial_\nu \tilde{A}_\mu\,.
\eeq
This replacement can be justified since $B$ is introduced with $dB=0$.
In this case,
we will find the same form of equations of motion with the one of the real $\mu$ theory given in the Lorentzian space-time.
However, in the present case,
we must notice that the solution $\tilde{\phi}$ is not simply a chemical potential but a combination of the chemical
potential and $\alpha$ as found from (\ref{Bfield}). 
This fact implies that we can obtain the solutions with $\tilde{\phi}$ for $\mu_I$ from the one of the real $\mu$ by a replacement,
$\mu/T\to i(\mu_I/T-\alpha)$. 

\vspace{.3cm}
As for the action $S_{F_{(4)}}$, this is necessary to study the potential of $\alpha$. 
Its explicit form and an effective potential of the B field is given in 2.2. 

\subsection{Bulk solutions}

As mentioned above, neglecting $\tilde{\phi}$, we can obtain solutions with the imaginary chemical potential $\mu_I$.
We give three solutions dual to the ground states of the pure YM fields, and they are compared.
Two of them are the solutions of $L_\mathrm{Gravity}$ only, then they are independent of $\mu$. 
The third solution is constructed by considering the back-reaction from $\tilde{F}^2$. 

\vspace{.3cm}
\noindent (1) AdS soliton solution: This represents the low temperature confinement phase, and
it is given as 
\beq\label{Soliton}
   ds^2=r^2(\delta_{\mu\nu}dx^{\mu}dx^{\nu}+f(r)dw^2)+{dr^2\over r^2f(r)}\, ,
\eeq
where 
\beq
 f(r)=1-\left({r_0\over r}\right)^5\, , \quad r_0={2\over 5R_w}\, ,
\eeq
and $2\pi R_w$ denotes the compactified length of $w$.  

\vspace{.3cm}
\noindent (2) The AdS-Schwarzschild solution: This solution corresponds to the high temperature deconfinement phase,
\beq\label{Sch}
   ds^2=r^2(fdt^2+\Sigma_i^3 (dx^i)^2 +dw^2)+{dr^2\over r^2f(r)}\, ,
\eeq
where 
\beq
 f(r)=1-\left({r_0\over r}\right)^5\, , \quad r_0={2\over 5R_w}\, ,
\eeq

\vspace{.3cm}
\noindent (3) Reisner-Nordstrom (RN): In this case, the back-reaction of flavor is taken into. It represents
the high temperature deconfinement phase.

The background of RN is given as the solution of the following action
\beq\label{B-F}
S_{G} = \int d^{6} x \sqrt{-g} \left\{{1\over 2\kappa_6^2}\left( {\cal R} + {20 \over L^2}\right)- {1 \over 4} \tilde{F}^2 \right\}\, ,
\eeq
which includes the flavor part. 
We get the following RN solution,
\beq\label{RN}
   ds^2=r^2(gdt^2+\Sigma_i^3 (dx^i)^2 +dw^2)+{dr^2\over r^2g(r)}\, ,
\eeq
\beq
 g=1-(1-{3\tilde{\mu}^2\over 8 r_+^2})\left({r_+\over r}\right)^5-{3\tilde{\mu}^2r_+^6\over 8r^8}
\label{RN2}
\eeq
\beq\label{RN3}
  \tilde{A}_0=\tilde{\phi}=\tilde{\mu}\left(1-{r_+^3\over r^3}\right)
\eeq
Here $r_+$ denotes the horizon of the charged black hole, and the temperature is given as
\beq\label{T-RN}
  T={1\over 4\pi}\left(5r_++{9\tilde{\mu}^2\over 8r_+}\right)\, .
\eeq
Here, $\tilde{\mu}$ is defined by
\beq\label{tildmu}
   -{ \tilde{\mu}\over T}= \int_{D_2} \tilde{F} = \int_{D_2}\left(F+ {B\over 2\pi\alpha'}\right)  =  \alpha-{\mu_I \over T} \, .
\eeq

\vspace{.3cm}
The action densities for these solutions, (1) AdS-Soliton (2) AdS-Schwartzchild and (3) RN, are given as
\bea
 S_{1}/V_3 &=& -r_0^5v_2 =-r_0^5{4\pi\over 5r_0}{1\over T}\, \\
 S_{2}/V_3 &=& -r_0^5v_2=-r_0^5\left({4\pi\over 5r_0}\right)^2\, \\
 S_{3}/V_3 &=& -r_+^5\left(1-{3\tilde{\mu}^2\over 8r_+^2}\right)v_2= -r_+^5\left(1-{3\tilde{\mu}^2\over 8r_+^2}\right){4\pi\over 5r_0}{1\over T}\, 
 \label{RN-potential}
\eea
where $v_2=\int_0^{\beta} d\tau\int dw$, $V_3=\int dxdydz$. 
In $\mu_I -T$ plane, we can find the phase diagram by comparing the above three actions. 
We find that the phase of the solution (2) is not realized when the solution (3) is added. 

\vspace{.5cm}
\subsection{Potential of Kalb-Ramond field}

Consider the
Kalb-Ramond
 field 
$B$, which is introduced in terms of $\alpha$ 
which is defined by (\ref{B-alpha}).
The present bottom up model would be related to D4/D8 model of type IIA string. Then
the bulk action, which could provide the potential of $\alpha$, might be given as
\be\label{Pot-1}
 S_{F_{(4)}}=-{1\over 2\kappa_6^2}\left(\int d^{6}x\,\sqrt{g}\,
\frac{1}{12} F_{(4)}^2 - \int B\wedge F_{(4)}\right ).
\ee
In the RN background, (\ref{RN})-(\ref{T-RN}), this action is estimated for a constant field $F_{123w}$ and $\alpha$.
We obtain
\be\label{Pot-2}
 S_{F_{(4)}}=-{V_4\over 2\kappa_6^2}\left({\beta\over 6r_+^3} F_{123w}^2-\alpha F_{123w} \right ).
\ee
where $V_4=\int d^3x~ dw=\beta V_3$, $V_3=\int dx^3$, and $r_+$ is used as the lower limit of the integration of $r$ as
$\int_{r_+}^{\infty} dr$. It should be replaced by $r_0$ in the case of solutions (1) and (2).
{Solving the equation of motion of $F_{123w}$, we find the solution as $F_{123w}=3 r_+^3 \alpha/\beta$}. Then the potential is obtained as
\be\label{Pot-3}
 S_{F_{(4)}}/V_3 \equiv 
V_A={1\over 2\kappa_6^2} {3r_+^3\over 2} \alpha^2 .
\ee
Due to the gauge symmetry of the dual SYM theory, the above result should be written as
\be\label{Pot-4}
     V_A=\min_{n\in {\mathbb Z}}{1\over 2\kappa_6^2}{3r_+^3\over 2} (\alpha - {2\pi n\over N_c})^2.
\ee

\section{Roberge-Weiss transitions}
\subsection {Probe approximation}

In the probe approximation, the gauge term $\tilde{L}_\mathrm{CSC}$ is treated as the probe for the background given by the gravitational
part. In this case, 
the background actions are given by $ S_{1}$ and $S_{2}$, and we find the critical line of confinement/deconfinemen by. 
comparing the two bulk actions. Then, the critical line is found as
\beq
   T={5r_0\over 4\pi}\, .
\eeq
This is independent of $\mu_I$, then the critical line is common to the case of real $\mu$.
The probe part defined by (\ref{probe-L}) is solved
under these backgrounds. 
The equations of motion of $\tilde{\phi}$ is given by the ansatz, $\tilde{A}=\tilde{A}_{\mu}dx^{\mu}=\tilde{\phi}(r)\,dt$. 

In the confinement phase, the background is given by the solution (1), and the equation for $\tilde{\phi}$ is given as
\begin{equation}
         \tilde{\phi}''+\left(\frac{4}{r}+\frac{f'}{f}\right)\tilde{\phi}' =0\;.
\label{eq2}
\end{equation}
We find that the allowed solution of this equation is 
$\tilde{\phi}=$const.. This solution gives 
no contribution to the free energy. On the other hand, the Kalb-Ramond field has no meaning in the confinement phase.  So there is no
new phase transition in this phase.

\vspace{.3cm}
An interesting phenomenon is observed in the deconfinement phase with the
background of the solution (2). In this case, we have the equation,
\begin{equation}
\tilde{\phi}''+\frac{4}{r}\tilde{\phi}'=0\;. \label{eq22}
\end{equation}
This equation is solved as 
\be\label{probe2}
  \tilde{\phi}=\tilde{\mu} \left( 1-{r_0^3\over r^3}\right)\, .
\ee
This solution provides a non-trivial contribution to the free energy as shown below. 
The probe action is given as
\bea
  S_{CS}^{E} &=& -\int dx^6 \sqrt{-g} \left(- {1 \over 4} \tilde{F}^2 \right)\\
      &=&\int dx^5 \int_{r_0}^{\infty} dr {r^4\over 2}(\tilde{\phi}')^2 \\
       &=&\int dx^3 V_f\,,
\eea
where
\be
   S_{\rm CS}^{\rm E}/V_3 = V_f={3\over 2}\left({4\pi\over 5}\right)^3 T \tilde{\mu}^2\, .
\ee
Then $\tilde{\mu}$ is replaced to $\mu_I$ and $\alpha$ by (\ref{tildmu}), and we obtain
\be
   V_{\rm f}
                   ={3\over 2}\left({4\pi T\over 5}\right)^3  (\alpha-\mu_I/T)^2\, .
\ee

\begin{figure}[htbp]
\vspace{.3cm}
\begin{center}
\includegraphics[width=7.0cm,height=7cm]{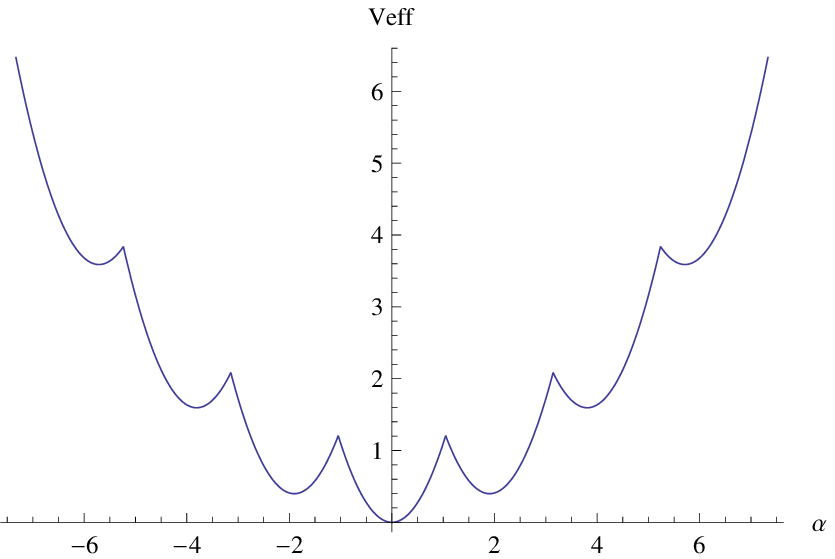}
\includegraphics[width=7.0cm,height=7cm]{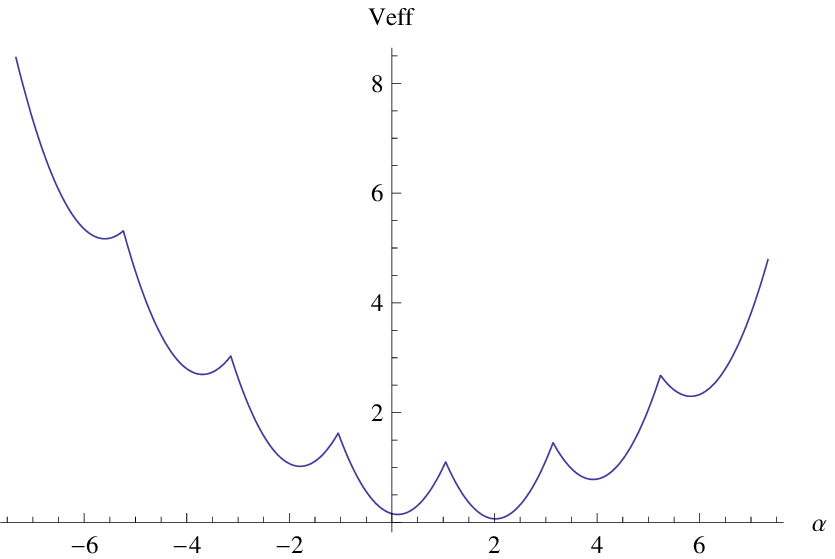}
\caption{$V_{\rm eff}$ for  $\mu_I/T=0$ (left) and $\mu_I/T=0.6a$ (right). 
}
\label{Phase-diagram-3-2}
\end{center}
\end{figure}

\begin{figure}[htbp]
\vspace{.3cm}
\begin{center}
\includegraphics[width=10.0cm,height=7cm]{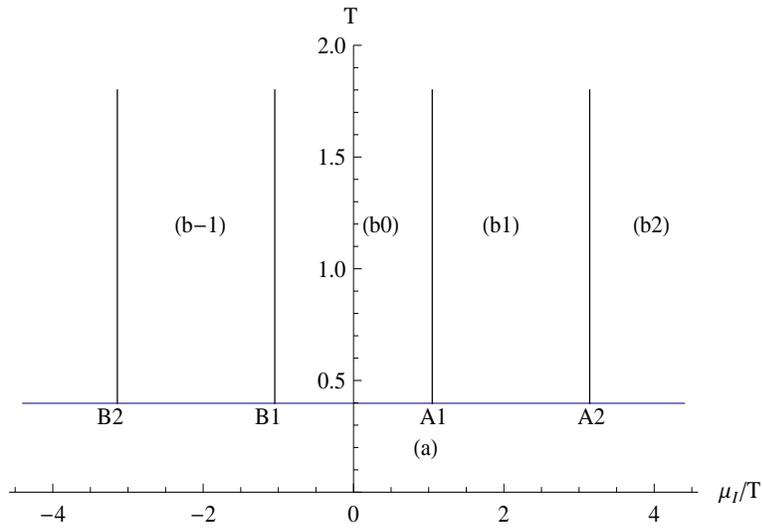}
\caption{Phase diagram 
for probe approximation. The horizontal critical line separates the confinement and deconfinement phases.
In the large $T$ deconfinement phase, the RW transitions are shown by the vertical critical lines. 
The points $A_1 \sim B_2$ represent tri-critical points.}
\label{Phase-diagram-3-3}
\end{center}
\end{figure}

\vspace{.3cm}


Now, from this  $V_{\rm f}$ and (\ref{Pot-4}) where $r_+$ is replaced by $r_0(=4\pi T/5)$, we find 

\be\label{Pot-5}
    V_{\rm eff} = \,V_A+V_f(\alpha)\,=\,
     \min_{n\in {\mathbb Z}}{1\over 2\kappa_6^2}{3\over 2}\left({4\pi T\over 5}\right)^3 (\alpha - {2\pi n\over N_c})^2
    + {3\over 2}\left({4\pi T\over 5}\right)^3  (\alpha-\mu_I/T)^2\, .
\ee

From this effective potential we could see the Roberge-Weiss transition for the state
defined by the value of $\alpha$. An example of this transition is read from 
the Fig.,  \ref{Phase-diagram-3-2} in which  we can see the transition from $\langle\alpha\rangle=0$ to 
$\langle\alpha\rangle=2\pi/3$ vacuum state, namely from the phase (b0) to (b1) in the Fig. \ref{Phase-diagram-3-3}. 
The resultant phase-diagram obtained from the above $V_{\rm eff}$ is shown
in the Fig. \ref{Phase-diagram-3-3}.
{Here, the relative ratio of the probe term and the B term $V_A$ is set by the relation $1/2\kappa^2_6=10$ 
for the simplicity.}

\begin{figure}[htbp]
\vspace{.3cm}
\begin{center}
\includegraphics[width=10.0cm,height=7cm]{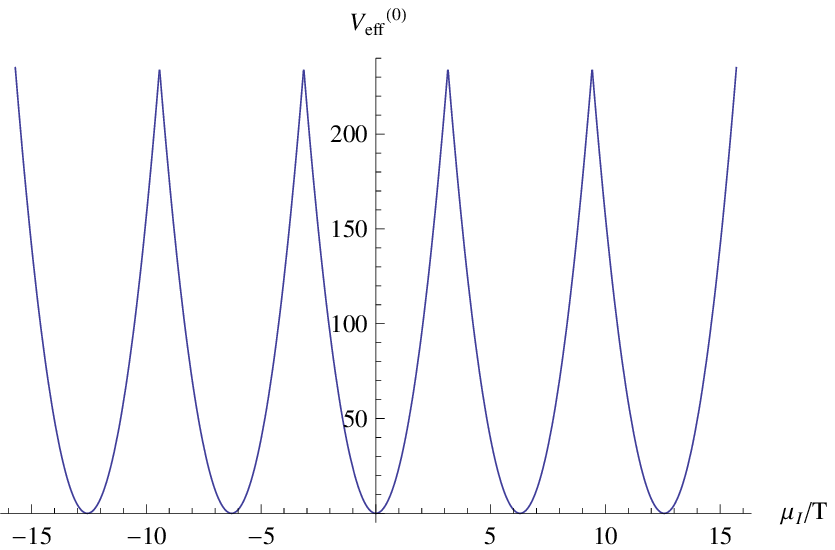}
\caption{$V_{\rm eff}^{(0)}$ }
\label{Potential-2}
\end{center}
\end{figure}

\vspace{.5cm}
Finally, we give an effective potential under the quenched approximation of the gauge field configurations
which provide the real Polyakov loop. This potential is obtained from (\ref{Pot-5}) by considering the functions at $\alpha=2\pi n$
where ${n\in {\mathbb Z}}$, and it is found by picking up the minimum parts,
\beq
    V_{\rm eff}^{(0)} = \min_{n\in {\mathbb Z}}  {3\over 2}\left({4\pi T\over 5}\right)^3  (2\pi n-\mu_I/T)^2\, .
\eeq
This potential has the period $2\pi$ with respect to $\mu_I/T$ as expected, and it is shown in the Fig. \ref{Potential-2}.
This period could be understood from the phase of the boundary condition imposed on the fundamental fermions of the
theory. In the present article, this potential is not used, however, this periodicity is seen for example in the calculation of the
chiral condensate in the gauge configurations of the real Polyakov loops \cite{Bilgici:2008qy,Bilgici:2009tx}. On this point,
the discussion is given more in the Sec. 5.

\vspace{.3cm}
\subsection{Back-reacted case}

When the back-reaction of the flavor part is taken into account,
the deconfinement background is replaced by the RN solution (3) since $S_3<S_2$.
The action $S_3$ is given in 
(\ref{RN-potential}). We notice the $\tilde{\mu}$ dependence
of $S_3$ is also comes from $r_+$. Actually, by using (\ref{T-RN}),
$r_+$ is written as 
\beq\label{r-p}
  r_+={2\pi T\over 5}\left(1+\sqrt{1-{45\tilde{\mu}^2\over 8}/(2\pi T)^2}\right)\, .
\eeq
This solution is useful for $|\tilde{\mu}|/T<\sqrt{32\pi^2/45}(=2.65)$ due to the reality of $r_+$.
For $|\tilde{\mu}|/T>\sqrt{32\pi^2/45}$, the system becomes unstable and decays to the stable confinement phase
expressed by the solution (1), the AdS-Soliton background.

\begin{figure}[htbp]
\vspace{.3cm}
\begin{center}
\includegraphics[width=10.0cm,height=7cm]{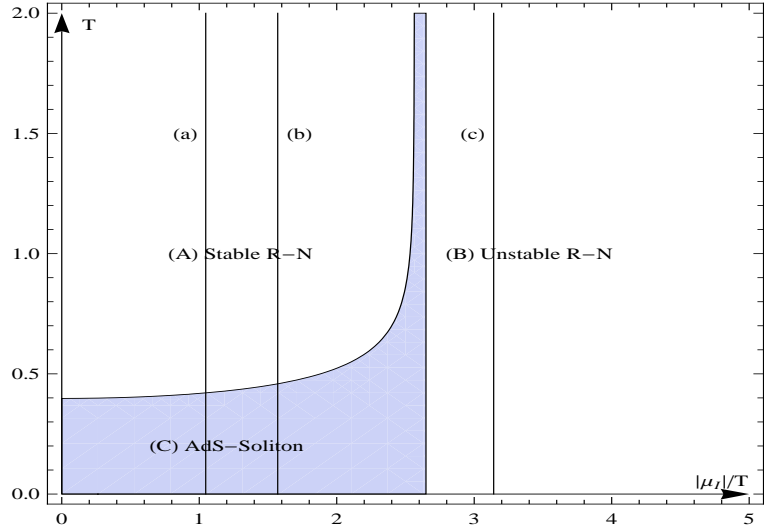}
\caption{Phase diagram of the back reacted case for $r_0=1$ and $\alpha=0$.
The lines (a), (b) and (c) represent  $|\mu_I|/T=\pi/3, \pi/2,$ and $\pi$ 
respectively.
}
\label{Phase-diagram-2}
\end{center}
\end{figure}
To observe the RW transitions, it is helpful to see
the phase diagram for $\alpha=0$, namely for $\tilde{\mu}=\mu_I$. 
This diagram is obtained by comparing $S_3$ with $S_1$ for $\alpha=0$, and it is shown 
in the Fig. \ref{Phase-diagram-2}.
The unstable region of solution (3) mentioned above
is shown by the region (B) in this figure for $\alpha=0$, and 
the regions (A) for Reisner-Nordstrom deconfinement phase and (C) for AdS soliton confinement phase are shown. 
Since the region (C) must be replaced by the confinement phase of (B), then
the deconfinement region 
(A) is restricted to the definite region of $|\mu_I|/T$ , $|\mu_I|/T\leq\sqrt{32\pi^2/45}$, as mentioned above. 
This restriction comes from the back-reaction.

\vspace{.3cm}
Now we study the RW transitions in the deconfinement phase of region (A) by reviving $\alpha$. 
In the present case, 
the effective potential is given as
\be
V_{\rm eff} = \,V_A+V_f^{\rm RN}\,=\,\min_{n\in {\mathbb Z}}{1\over 2\kappa_6^2}{3r_+^3\over 2} (\alpha - {2\pi n\over N_c})^2
 +  V_f^{\rm RN}
\label{genericmass-1}
\ee
The first term $V_A$ is obtained in (\ref{Pot-4}) for the background of RN, and the second term
is given as 
\beq\label{Vf}
S^{\rm RN} ={1\over 2\kappa_6^2} \int d^{6} x \sqrt{-g} \left\{ {\cal R} + {20 \over L^2}- {1 \over 4} \tilde{F}^2 \right\}=\int dx^3 ~V_f^{\rm RN}\, .
\eeq
Using (\ref{RN-potential}), we find
\be
   V_f^{\rm RN} = -{1\over 2\kappa_6^2}r_+^5\left(1-{3\tilde{\mu}^2\over 8r_+^2}\right){4\pi\over 5r_0}{1\over T}\, ,
\ee
This potential is the part dual to the combined system of SYM fields and flavor fermions with an imaginary chemical potential
$\mu_I$. 

\vspace{.3cm}
Here, we notice that also $V_A$ depends on $\tilde{\mu}$ through $r_+$. 
This fact can be interpreted as a kind of the back-reaction to the Kalb-Ramond potential from the flavor fermions.
In order to 
understand this back-reaction, we restrict the region of $|\mu_I|$
to the region of small $|\mu_I/T-\alpha|$. 
Then we can expand $V_{\rm eff}$ in the series of 
$-(\mu_I/T-\alpha)^2$. 
The expanded potential is retained
up to the order of  $|\mu_I-\alpha/\beta|^2$, and we obtain
\bea
      V_A &=&\, \min_{n\in {\mathbb Z}}{1\over 2\kappa_6^2}
  \left({64\pi^3\over 125}-{27\pi\over 50} \left(\alpha-{\mu_I\over T} \right)^2\right)  T^3 (\alpha - {2\pi n\over N_c})^2+ \cdots  
                              \label{VA1}  \\
    V_f^{\rm RN} &= &{1\over 2\kappa_6^2} \left(- {1024\pi^5\over 3124}+{96\pi^3\over 125}\left(\alpha-\frac{\mu_I}{T}\right)^2 \right) T^3
        +\cdots\,.
\label{genericmass-2}
\eea
We find that this result is almost equal to the probe approximation
except the point that the coefficient of $V_A$ is slightly modified. In fact, we can see a similar
behavior of the potential
to the one of the probe approximation. Then we could find the expected RW transitions in this case also.
Furthermore, the qualitative behaviors of the potential are maintained even if
the full form of potential is used.  
So we show here the RW transitions in terms of the full form of potential.

\begin{figure}[htbp]
\vspace{.3cm}
\begin{center}
\includegraphics[width=7.0cm,height=7cm]{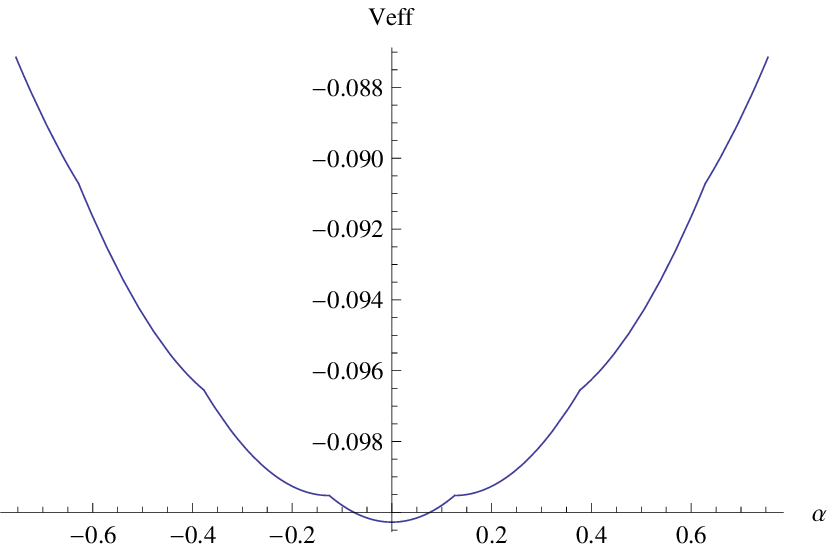}
\includegraphics[width=7.0cm,height=7cm]{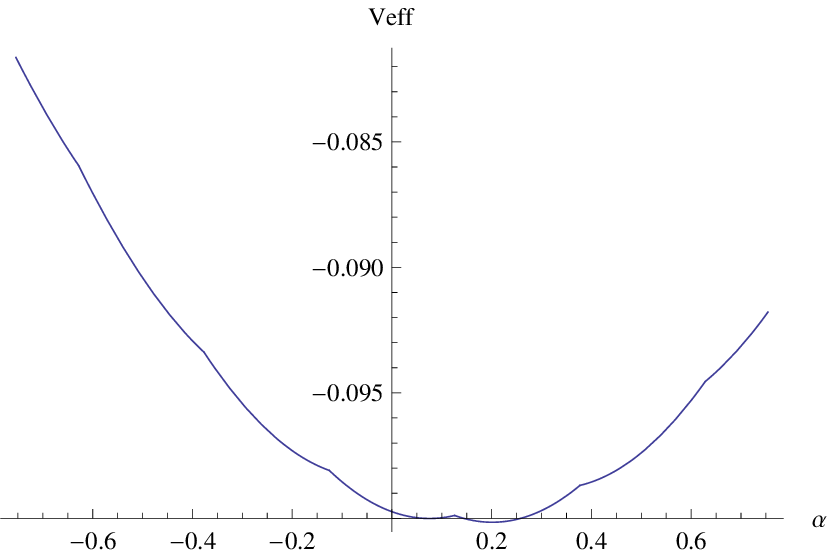}
\caption{Full form of $V_{\rm eff}$ for RN back-reacted case. Left is for $\mu_I/T=0$, and right is for $\mu_I/T=0.6a$ with
the period of $a=2\pi/25$.}
\label{Phase-diagram-6}
\end{center}
\end{figure}

\vspace{.3cm}
\begin{figure}[htbp]
\vspace{.3cm}
\begin{center}
\includegraphics[width=10.0cm,height=7cm]{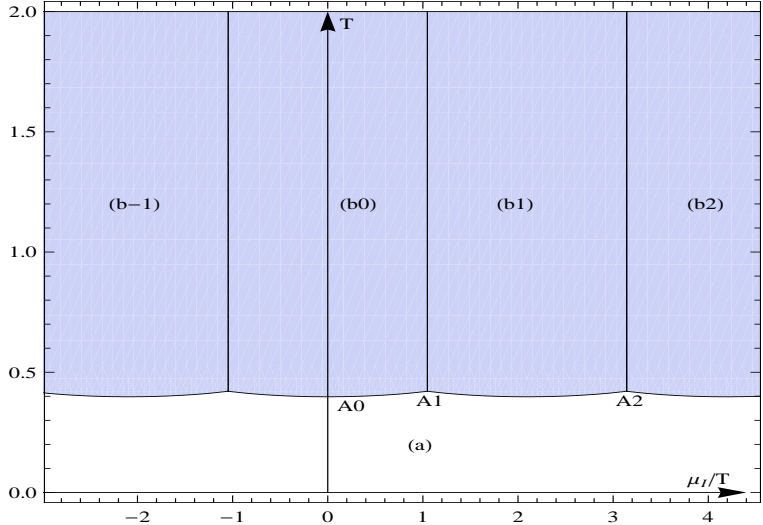}
\caption{Phase diagram for RW transitions under the back reacted background solution for $N_c=3$.
In the RN deconfinement phase, the phases are separated by the vertical critical-lines to
the regions, (b-1) $\sim$ (b2) for  $\langle\alpha\rangle=-2\pi/3,~0,~2\pi/3, ~4\pi/3$. 
The area (a) corresponds the AdS soliton confinement phase. 
The points 
$A_1$ and $A_2$ represent the tri-critical points.}
\label{Phase-diagram-4}
\end{center}
\end{figure}

In the Fig. \ref{Phase-diagram-6}, the effective potential with two values of $\mu_I/T$ are shown to see the RW transition from $\langle\alpha\rangle =0$ 
to $\langle\alpha\rangle =a$ with the period $a$, which should be set as $2\pi/N_c$. It is easy to find
other periodic transitions. 
The resultant phase diagram with the RW transitions is shown in the Fig. \ref{Phase-diagram-4} .

Here, we should notice that the periodicity of $\mu_I/T$ in $V_{\rm eff}$ implies $|\mu_I/T|<\pi/N_c$.
This and the constaint, $|\mu_I|/T\leq\sqrt{32\pi^2/45}$, given above leads to the constraint,
\beq
    N_c>1.18.
\eeq
In spite of the fact that $N_c$ must be a large integer to justify the holographic approach, we are
allowed to use the holography up to $N_c=2$ at this stage.

\section{Comparison with QCD near $\mu=0$}

As shown above, the $\mu$ dependent critical line obtained for real $\mu$ can be 
continued to the imaginary $\mu$ region and used there.
The form of
the critical line near $\mu=0$ is given as \cite{EN},
\beq\label{smallmu}
   {T\over T_0}=1-a \left({\mu \over T_0}\right)^2+\cdots\, ,
\eeq 
where $T_0$ denotes the critical temperature at $\mu=0$, and $a$ is a dimensionless constant, which depends 
on the parameters of the theory. It has been obtained also in the lattice QCD for $\mu^2<0$
without bothering with the sign problem. Then it is meaningful to compare the result obtained in lattice QCD
with the holographic result given here. 

\vspace{.3cm}  

For the probe approximation, we find $a=0$ since the critical line is independent of $\mu$. 
So we consider
the RN background case, where the $\mu$-dependent critical curve 
is obtained from the equation
$S_1=S_3$. 
And we find
\beq\label{small-mu-b}
   a= {15 \over 32\pi^2 }= 0.0475 \, .
\eeq

\vspace{.3cm}  
In the lattice QCD,
the coefficient is obtained by the form
\beq\label{alatt}
  a=\kappa N_c^2\, .
\eeq
For confinement transition,  we find many simulation results. We pick up several examples, to compare with the 
holographic result, where $N_c$ is however assumed to be large. The examples of 
estimations of $\kappa$ from lattice QCD data in {the $2+1$ and $2+1+1$ flavor systems}; examples are
{$0.0066 \pm 20$~\cite{Endrodi:2011gv},}
$0.013 \pm 0.003$~\cite{Bonati:2014rfa}, $0.0135 \pm 0.002$~\cite{Bonati:2015bha}, $0.0149 \pm 0.0021$~\cite{Bellwied:2015rza}, $0.020 \pm 0.004$~\cite{Cea:2015cya}. 
Comparing this with (\ref{small-mu-b}) and using (\ref{alatt}), we find $N_c=1.76$ {for $\kappa \sim 0.0153$}.
This is consistent wth the result $N_c\geq 1.2$ obtained in our back reacted case.

\vspace{.3cm}
In addition, there is the estimation of $\kappa$ by using the Polyakov-loop extended Nambu--Jona-Lasinio (NJL)
model with the mean-field approximation as
{$\kappa_\chi = 0.017 \pm 0.001$~\cite{Kashiwa:2017yvy}};
it should be noted that this value is estimated from the iso-spin chemical potential but $\kappa$ should be exactly same in both cases at least in the mean-field approximation.  
{In the case of the PNJL model, there is the estimation of $\kappa$ for the confinement-deconfinement crossover line as $0.004 \pm 0.001$ and $0.003 \pm 0.001$; the former one is evaluated from the Polyakov-loop and the later one does the quark number holonomy~\cite{Kashiwa:2016vrl}.}


\section{More about the periodicity}

In previous section, we discuss possible connections between lattice QCD simulation near $\mu^2=0$.
In this section, we discuss deeper properties of the periodicity appearing in QCD.
In full QCD, we should have the RW periodicity as explained and demonstrated above.
It is, however, well known that we should have the $2 \pi$ periodicity for $\mu_I/T=\theta$ instead of the
Roberge-Weiss (RW) periodicity in the lattice QCD simulation when we fix
gauge configurations at $\mu=0$ or the pure gauge limit where dynamical quarks are not taken into account;
later one is corresponding to the quenched limit.
In this case, any quantities such as the pressure, the entropy density and the cumulant loses the RW periodicity and thus the minimal period becomes $2\pi$ because the grand canonical partition function does not have the RW periodicity.
It should be noted that we need the Polyakov-loop phase-flip to
consider the RW periodicity in the limits as discussed in
Ref.~\cite{Doi:2017dyc,Doi:2017jad}.

The $2\pi$ periodicity has been used in the calculation of the dual quark
condensation.
Actual definition of the dual quark condensation is given by
\beq
 \Sigma^{(n)} = \int_0^{\textcolor{red}{2}\pi} \frac{d \phi}{2 \pi} \sigma(\phi) e^{i n \theta} d \phi
\eeq
where $\sigma(\phi)$ is the chiral condensate with the phase of the boundary
condition, $0 \le \phi \le 2 \pi$, which is related with the dimensionless imaginary
chemical potential as
\beq 
\phi = \theta + \pi.
\eeq 
In the heavy quark mass regime, it has the clear relation with the Polyakov-loop by using the Dirac
mode expansion~\cite{Bilgici:2008qy,Bilgici:2009tx}.
Since the dual quark condensate is calculated from the chiral condensate,
this quantity may bridge the chiral and the Polyakov-loop dynamics in QCD.
Details of the dual quark condensate have been discussed in the lattice
QCD simulation~\cite{Bilgici:2008qy,Bilgici:2009tx}, the
Dyson-Schwinger equation~\cite{Fischer:2009wc}, QCD effective
models~\cite{Kashiwa:2009ki,Xu:2011pz,Benic:2013zaa}.
In principle, we can investigate the Polyakov-loop behavior at $\mu=0$ from
the $\phi$-dependent chiral condensate even if calculations of the
Polyakov-loop is difficult or impossible.

In the strict probe limit of the holographic model, we also face the same
situation of the lattice QCD simulation in the quenched limit;
we only has the $2\pi$ trivial periodicity.
For the consistency between the holographic model and the lattice QCD
simulation in both limits, we should reproduce the $2 \pi$ periodicity in
addition to the RW periodicity.
Also, it is good to calculate the dual quark
condensate in the holographic model to discuss the relation between the
chiral condensate and the Polyakov-loop. In addition, if we can
understand how to control the boundary condition in the holographic
model, we will contact with the ${\cal Z}_\mathrm{N_c}$ twisted QCD; it
is an interesting QCD like theory in the viewpoint of the sign problem
appearing in the lattice simulation.
{One possibility to introduce the $2\pi$ trivial periodicity in the probe limit is
imposing $2\pi$ periodic form of $\theta$ in the mapping of $\mu$ from $A_0$: 
we need the special care for the $2\pi$ periodicity issue of the
dimensionless imaginary chemical potential.



\vspace{1cm}
\section{Summary}

We have studied here the phase structure and phase transition behaviors ranging from
real chemical potential to imaginary one, using a bottom-up holographic model that was introduced 
to investigate color superconductivity in QCD. 
From general framework of QCD, one knows that
the QCD partition function possesses a certain periodicity, the Roberge-Weiss (RW) periodicity,
at the imaginary chemical potential region. Our interest was to see how the analytic continuation of 
the chemical potential works.
To this end, we have computed the effective potential of the model by including the Kalb-Ramond field
in the bulk. Unlike the previous studies based on top-down approach, there is an advantage of our bottom-up approach that one can evaluate the effect of back reaction. As the result, we have observed the RW periodicity as well as the $2\pi$ periodicity appropriately. 
We have further investigated the behavior of the critical line near $\mu=0$ and tried to see the validity
of our analysis. Our results have been compared with those obtained from lattice QCD
and effective models such as Polyakov-loop extended NJL model

\vspace{.3cm}
\section*{Acknowledgments}
The authors would like to thank helpful discussion with Takeshi Morita.
{K.K.is supported by the Grants-in-Aid for Scientific Research from JSPS (No. 18K03618).}

\newpage
\vspace{.5cm}

\appendix

\noindent{\bf\Large Appendix}


\vspace{.5cm}
\def\theequation{\thesection.\arabic{equation}}
\setcounter{equation}{0}

\section{Roberge-Weiss periodicity in the operator formalism}

\setlength{\arraycolsep}{2pt}
\renewcommand{\bra}[1]{\left.\langle#1\right|}
\renewcommand{\ket}[1]{\left|#1\rangle\right.}
\newcommand{\bracket}[2]{\langle#1|#2\rangle}
\renewcommand{\tr}{\mathrm{tr}}

Since the Roberge-Weiss periodicity must be independent of the gauge fixing
condition except the global topology of gauge configuration,
the mechanism of the periodicity seems to result strongly from the
quark side.
In this appendix, we present a brief demonstration of the mechanism
in the operator formalism.
In the path-integral representation,
the periodicity is related to the boundary condition along the imaginary time.
Therefore, we omit the spatial degrees of freedom in the following
discussions.

Let us begin with a single fermion system. The coherent states are defined by
\begin{equation}
\ket{\xi}=(1-\xi a^\dagger)\ket{0}=\ket{0}-\xi\ket{1}\ ,\qquad
\bra{\xi}=\frac{1}{i}\bra{0}(\xi-a)=\frac{1}{i}\left(\bra{0}\xi-\bra{1}\right)
\ .
\end{equation}
Using the integration rule $\int\!d\xi\,\xi=i$,
the trace of an operator $\mathcal{O}$ is calculated as
\begin{equation}
\int\!\!d\xi\bra{\xi}\mathcal{O}\ket{-\xi}
=\bra{0}\mathcal{O}\ket{0}+\bra{1}\mathcal{O}\ket{1}
=\tr\,\mathcal{O}\ .
\end{equation}
This leads to the anti-periodic boundary condition
in a path-integral representation~\cite{OhKs}.

In a particle-antiparticle system, the coherent states which satisfy
\begin{equation}
a\ket{\xi,\bar{\xi}}=\xi\ket{\xi,\bar{\xi}}\ ,~~
b\ket{\xi,\bar{\xi}}=\bar{\xi}\ket{\xi,\bar{\xi}}\ ,~~
\bra{\xi,\bar{\xi}}a=\bra{\xi,\bar{\xi}}\xi\ ,~~
\bra{\xi,\bar{\xi}}b=\bra{\xi,\bar{\xi}}\bar{\xi}
\end{equation}
are constructed as follows:
\begin{eqnarray}
\ket{\xi,\bar{\xi}}
&=&(1-\xi a^\dagger)(1-\bar{\xi}b^\dagger)\ket{0,0}
=\ket{0,0}-\xi\ket{1,0}-\bar{\xi}\ket{0,1}+\bar{\xi}\xi\ket{1,1}\ ,\\
\bra{\xi,\bar{\xi}}
&=&\frac{1}{i^2}\bra{0,0}(\bar{\xi}-b)(\xi-a)
=\frac{1}{i^2}\left(
\bra{0,0}\bar{\xi}\xi+\bra{1,0}\bar{\xi}-\bra{0,1}\xi+\bra{1,1}\right)\ .
\end{eqnarray}
Corresponding to boundary conditions as
$\psi(\beta)=-\sigma\psi(0)$ and $\bar{\psi}(\beta)=-\sigma^*\bar{\psi(0)}$
in the path-integral representation,
the calculation of the trace is modified as
\begin{eqnarray}
\tr_\sigma\,\mathcal{O}
&=&\int\!\!d\xi\,d\bar{\xi}
\bra{\xi,\bar{\xi}}\mathcal{O}\ket{-\sigma\xi,-\sigma^*\bar{\xi}}\\
&=&\bra{0,0}\mathcal{O}\ket{0,0}+\sigma\bra{1,0}\mathcal{O}\ket{1,0}
+\sigma^*\bra{0,1}\mathcal{O}\ket{0,1}+\bra{1,1}\mathcal{O}\ket{1,1}
\label{eq:modifiedtrace}
\end{eqnarray}
(This reduces to the normal trace in the case of $\sigma=1$.)

Note that
\begin{equation}
\ket{-\sigma\xi,-\sigma^*\bar{\xi}}
=\mathrm{e}^{i\gamma(a^\dagger a-b^\dagger b)}\ket{-\xi,-\bar{\xi}}
\qquad(\sigma=\mathrm{e}^{i\gamma})\ ,
\end{equation}
then we obtain
\begin{equation}
\tr_\sigma\,\mathcal{O}
=\tr\,(\mathcal{O}\,\mathrm{e}^{i\gamma(a^\dagger a-b^\dagger b)})\ .
\label{eq:trace_relation}
\end{equation}

Now, the grand potential $Z(\beta,\mu)$ of the $SU(N)$ local gauge theory
is invariant under the $Z_N$ transformation associated with
the $Z_N$-twisted boundary condition at $\tau=\beta$~\cite{rw}.
This is expressed, in the operator formalism, as
\begin{equation}
Z(\beta,\mu)=\sum_A\int\![d\xi d\bar{\xi}]\bra{A,\xi,\bar{\xi}}
\mathcal{O}\ket{A,-\sigma\xi,-\sigma^*\bar{\xi}}\ ,
\end{equation}
in which the label $A$ denotes a set of quantum numbers
other than the quark occupation and
\begin{eqnarray}
&&\bra{A,\xi,\bar{\xi}}
=\bra{A}\left(\mathop{\otimes}_{i=1}^N\bra{\xi_i,\bar{\xi}_i}\right)\ ,\qquad
[d\xi\,d\bar{\xi}]=d\xi_1d\bar{\xi}_1\cdots d\xi_Nd\bar{\xi}_N\ ,\\
&&\mathcal{O}
=\mathrm{e}^{-\beta(H-\mu\sum_{i=1}^N(a_i^\dagger a_i-b_i^\dagger b_i))}\ ,
\quad
\sigma=\mathrm{e}^{i\frac{2\pi k}{N}}\quad(k\in\{0,1,\ldots,N-1\})\ .
\end{eqnarray}
(For simplicity,
the degree of freedom associated to the Dirac spinor components is suppressed.)
Applying the relation (\ref{eq:trace_relation})
and taking account of the commutativity
among $H$, $a_i^\dagger a_i$ and $b_i^\dagger b_i$,
one immediately derives the RW periodicity:
\begin{eqnarray}
Z(\beta,\mu)
&=&\sum_A\int\![d\xi d\bar{\xi}]\bra{A,\xi,\bar{\xi}}
\mathrm{e}^{-\beta\left(
H-(\mu+i\frac{2\pi k}{\beta N})\sum_{i=1}^N(a_i^\dagger a_i-b_i^\dagger b_i
)\right)}\ket{A,-\xi,-\bar{\xi}}\\
&=&Z(\beta,\mu+i\frac{2\pi k}{\beta N})\ .
\end{eqnarray}



\end{document}